\documentclass[aps,prd,12point,twocolumn,nofootinbib,showpacs,superscriptaddress]{revtex4-1}
\def\theequation{\arabic{section}.\arabic{equation}}
\usepackage{psfrag}
\usepackage{subfigure}
\usepackage{color}
\usepackage{mathrsfs}
\usepackage{graphicx}
\usepackage{amssymb, bm}
\usepackage{amsmath, amsthm}
\usepackage{epstopdf}
\usepackage{hyperref}
\usepackage{enumerate}
\usepackage{longtable}

\newcommand{\be}{\begin{equation}}
\newcommand{\ee}{\end{equation}}

\begin{document}
\def\theequation{\arabic{section}.\arabic{equation}}

\title{Quasilocal mass and multipole expansion in scalar-tensor gravity}

\author{Valerio Faraoni}
\email{vfaraoni@ubishops.ca}
\affiliation{Department of Physics \& Astronomy, 
Bishop's University, 2600 College Street, Sherbrooke, Qu\'ebec, 
Canada J1M 1Z7 }

\author{Jeremy C\^ot\'e}
\email{jcote16@ubishops.ca}
\affiliation{Department of Physics \& Astronomy, 
Bishop's University, 2600 College Street, Sherbrooke, Qu\'ebec, 
Canada J1M 1Z7 }

\begin{abstract}

A generalization of the Hawking-Hayward quasilocal mass to scalar-tensor 
gravity is compared, {\em in vacuo} and for asymptotically flat stationary 
geometries, with a recent multipole expansion of the gravitational field. 
The quasilocal mass seen at spatial infinity coincides with the monopole 
term, lending credibility to this construct.

\end{abstract}


\maketitle

\section{Introduction}
\setcounter{equation}{0}

There is little doubt that general relativity (GR) is not the final theory 
of gravity, since it must break down at spacetime singularities, which 
must be somehow cured, and GR cannot be renormalized without 
altering 
it. All attempts to quantize gravity generate modifications to it in the 
form of higher order derivatives in the field equations or  curvature 
terms, non-local terms, or extra fields in the action 
\cite{Stelle}.\footnote{As an example, bosonic string theory (the simplest 
string 
theory) reduces to an $\omega=-1$ Brans-Dicke theory, and not to GR, in 
the 
low-energy limit \cite{bosonic}.} What is more, perhaps we are already 
observing modifications of GR in the present cosmic acceleration that 
the standard $\Lambda$CDM model of cosmology tries to fit into 
GR by invoking a completely {\em 
ad hoc} dark energy \cite{Amendolabook}. Maybe the acceleration is due to 
deviations from GR at large scales \cite{CCT}, an idea that has raised 
much interest and has led to the so-called $f({\cal R})$ gravity  
(see \cite{reviews, Amendolabook} for reviews). This class of 
theories is a subfamily of the broader scalar-tensor gravity that 
we consider here \cite{ST, DamourFarese}, and which generalizes the 
original Brans-Dicke proposal \cite{BransDicke}. There is currently a 
sustained theoretical and experimental effort to test gravity at multiple 
scales, in various regimes, and using various physical phenomena  
\cite{Willbook, tests, Padilla, Psaltis, Koyama, Euclid, smBHtests, 
gwtests}.

The notion of mass-energy in the relativistic physics of gravitating 
systems is complicated, because it must include gravitational energy.
However, due to the equivalence principle \cite{Willbook}, it is 
impossible to localize the energy of the gravitational field, which can 
be eliminated locally by changing to the frame of a freely falling 
observer. As the next best option, researchers have resorted to the use 
of quasilocal notions of energy, {\em i.e.}, integral quantities computed 
over  compact regions of space.  The Hawking-Hayward mass construct 
\cite{HawkingHayward}  
generalizes the Misner-Sharp-Hernandez mass---better known 
from fluid dynamics in spherical symmetry \cite{MSH}---and is universally 
adopted in black hole thermodynamics, but other constructs exist in the 
literature (see the review~\cite{Szabados}). 

Indeed, the quasilocal energy concept also appears in 
the first law of thermodynamics. Much literature has been 
devoted to horizon thermodynamics and to the thermodynamics 
of gravity and spacetime in GR and in alternative theories. 
Thus far, very few quasilocal mass prescriptions  have 
been given in scalar-tensor gravity, and they have important 
restrictions  \cite{previous, Cognola}. For example,  
a certain prescription applies only to $f({\cal R})$ gravity, or only 
to spherical  symmetry, and sometimes 
only to very special spherical 
spacetime geometries. These prescriptions do not agree with each other 
and have been obtained in the context of the   
thermodynamics of spacetime. However, 
the definitions of four other quantities appearing 
in the first law (temperature, entropy, work density, and 
heat supply vector) are not established beyond doubt, which 
introduces uncertainty in any definition of quasilocal mass 
obtained by {\em assuming} a certain form 
of the first law.  Additionally, the concept of temperature requires  
the study of quantum field theory in curved 
space, which is notoriously non-trivial, and where 
it is difficult to conclude calculations unambiguously (or to conclude them 
at all). 

While we remain agnostic on these approaches based on the thermodynamics 
of spacetime, we have proposed an 
extension of the Hawking-Hayward quasilocal mass to scalar-tensor gravity  
that bypasses these difficulties using arguments that are purely 
classical and independent of  
thermodynamics \cite{mySTmass}. The 
generalization of the Hawking-Hayward mass to 
scalar-tensor gravity is  
not restricted to $f({\cal R})$ gravity or to special geometries,  
nor does it require  spherical symmetry or asymptotic flatness (although 
we use the last two assumptions in our check of the validity 
of the quasilocal prescription in the present work).

This quasilocal mass generalization can be derived using two 
distinct approaches which give the same result {\em in vacuo},  
but differ slightly in the presence of matter \cite{mySTmass}. The first  
approach consists  of writing the scalar-tensor field equations as 
effective Einstein equations and using the geometric 
derivation of the Hawking-Hayward mass. It uses minimal assumptions and 
should be regarded as the most reliable prescription \cite{mySTmass}. The 
second approach relies on the Einstein frame formulation of scalar-tensor 
gravity, in which an omnipresent scalar field  couples 
nonminimally to matter, while the rest of the action looks like the 
Einsten-Hilbert one. Because of this feature, the theory is not strictly 
GR and this method produces an extra spurious   
factor $\phi^2 $  multiplying the 
matter contribution to the quasilocal 
mass (where $\phi$ is the Brans-Dicke scalar)  \cite{mySTmass}.

In the present article, we propose a partial check of our prescription for 
the quasilocal mass {\em in vacuo} \cite{mySTmass}. A minimal requirement 
for this prescription to  make sense is that, in  an 
asymptotically flat stationary geometry, the quasilocal mass reduces  
to the  coefficient of the monopole term in a multipole expansion of the 
gravitational field. Such a multipole 
formalism for scalar-tensor gravity has been introduced recently in 
Ref.~\cite{PappasSotiriou} for cylindrically symmetric and asymptotically 
flat systems. We show that the quasilocal mass computed at spatial 
infinity does indeed reduce to the coefficient of the lowest multipole, 
lending some confidence in the prescription of Ref.~\cite{mySTmass}.

Let us begin by reviewing basic scalar-tensor gravity. The (Jordan frame) 
scalar-tensor action is  \cite{BransDicke, ST}
\begin{eqnarray}
S_\text{ST}&=&\int d^4 x \sqrt{-g} \left\{ \left[ 
\frac{1}{16\pi} \left(  \phi {\cal R} 
-\frac{\omega(\phi)}{\phi} \, g^{ab} \nabla_a \phi \nabla_b 
\phi \right) \right. \right.\nonumber\\
&&\nonumber\\
&\, & \left. \left. -V(\phi) \right]  
+{\cal  L}^\text{(m)} \right\} \,,
\label{Jframeaction}
\end{eqnarray}
where ${\cal R}$ is the Ricci curvature of the spacetime 
metric $g_{ab}$ with determinant $g$, $\phi$ is the 
Brans-Dicke-like scalar field (roughly speaking, the 
inverse of the gravitational coupling strength which is 
varying in these theories \cite{BransDicke}), $V(\phi)$ is a scalar field 
potential, and ${\cal L}^\text{(m)}$ is the matter 
Lagrangian density.  The conformal rescaling of the metric 
\be \label{confo1}
g_{ab} \rightarrow \tilde{g}_{ab}=\Omega^2 \, g_{ab} \,, 
\;\;\;\;\;\; \Omega=\sqrt{\phi} 
\ee
and the scalar field redefinition $\phi \rightarrow 
\tilde{\phi}$ given by
\be \label{confo2}
d\tilde{\phi} =\sqrt{ \frac{2\omega(\phi)+3}{16\pi}} \, 
\frac{d\phi}{\phi}
\ee
transform the action to its Einstein frame form 
\begin{eqnarray}
S_\text{ST} &=& \int d^4 x \sqrt{-g}  \left[ 
\frac{\tilde{{\cal R}} }{16\pi} -\frac{1}{2} \, 
\tilde{g}^{ab} 
\nabla_a \tilde{\phi} \nabla_b \tilde{\phi}  
-U (\tilde{\phi}) \right. \nonumber\\
&&\nonumber\\
&\, &  \left. + \frac{{\cal L}^\text{(m)} }{
\phi^2} \right] \, .\label{Eframeaction}
\end{eqnarray}
This is formally the action of GR with a scalar field 
coupling minimally to the curvature (but nonminimally to 
matter), where the scalar field has canonical kinetic energy density and 
\be
U( \tilde{\phi})= \frac{ V (\phi)}{\phi^2} 
\Big|_{\phi=\phi \left( \tilde{\phi} \right) }  \,.
\ee 
(A tilde identifies Einstein frame quantities.)  Formally, the 
difference with GR is in the 
nonminimal (but universal) coupling of 
the Einstein frame scalar $\tilde{\phi}$ to 
(non-conformal) matter. This coupling becomes irrelevant {\it in 
vacuo}; otherwise, one can 
regard the scalar field $\tilde{\phi}$ and the matter 
fields described by ${\cal L}^\text{(m)}$ as mutually 
interacting forms of matter, with the condition that the scalar field 
$\tilde{\phi}$ is always present and cannot be switched 
off.

$f({\cal R})$ theories of gravity, widely used to explain the present 
acceleration of the universe without dark energy \cite{CCT, reviews}, 
are a  subclass of scalar-tensor theories with action
\be
S_{ f({\cal R}) } = \int d^4 x \sqrt{-g} \, f({\cal R}) + 
S_\text{(m)}
\ee
where $f({\cal R})$ is a nonlinear function of the Ricci 
scalar. By setting $\phi = f'({\cal R})$ and
\be
V(\phi)= \phi {\cal R}(\phi) -f\left( {\cal R}(\phi)  
\right) \,,
\ee
the action can be shown to be  equivalent to the 
scalar-tensor one \cite{reviews}
\be
S = \int d^4 x \, \frac{ \sqrt{-g}}{16\pi}  \left[ \phi {\cal 
R}-V(\phi) \right] +S_\text{(m)} \,,
\ee
which describes a Brans-Dicke theory with vanishing Brans-Dicke parameter  
$\omega$ and potential $V$ for the Brans-Dicke scalar 
$\phi$.

\section{Spherical symmetry}
\setcounter{equation}{0}

For simplicity, in our test of the scalar-tensor quasilocal prescription, 
we restrict to vacuum and spherical symmetry. With this symmetry, the 
Hawking-Hayward 
mass in GR reduces \cite{Haywardspherical} to the better known 
Misner-Sharp-Hernandez mass 
$\tilde{M}_\text{MSH}$ defined by \cite{MSH}
\be \label{MSHdefinition}
1-\frac{2\tilde{M}_\text{MSH}}{ \tilde{R}}=\tilde{g}^{ab} \nabla_a 
\tilde{R} \, \nabla_b \tilde{R} \,,
\ee
where $ \tilde{R}$ is the areal radius of the spherical GR spacetime. 

One of the two methods used in Ref.~\cite{mySTmass} to construct the 
scalar-tensor generalization of the Hawking-Hayward mass in the Jordan 
frame is by conformally 
transforming this mass to the Einstein frame and imposing that, in the Einstein frame, the 
usual GR expression holds and that this mass transforms as the 
Hawking-Hayward mass 
under a conformal mapping to the Jordan frame. Note that the second, and more 
reliable method, gives the same result {\it in vacuo}.  

Under the conformal rescaling $g_{ab} 
\rightarrow  \tilde{g}_{ab}=\Omega^2 g_{ab}$, the 
Misner-Sharp-Hernandez mass transforms according to 
\cite{HHconfo, FaraoniVitagliano2014}
\be
\tilde{M}_\text{MSH} = \Omega M_\text{MSH} 
-\frac{R^3}{2\Omega} \, g^{ab} \nabla_a \Omega\nabla_b 
\Omega -R^2 g^{ab} \nabla_a \Omega\nabla_b R \,.  \label{MSHtransformation}
\ee
Specializing to the particular conformal 
transformation~(\ref{confo1}) and (\ref{confo2}), 
Eq.~(\ref{MSHtransformation}) then yields the Jordan frame mass
\be
M_\text{JF}=\frac{\tilde{M}_\text{MSH} }{\sqrt{\phi}} 
+\frac{R^3}{8\phi^2} \, g^{ab}\nabla_a\phi \nabla_b \phi 
+\frac{R^2}{2\phi}  \, g^{ab}\nabla_a\phi \nabla_b R \,,
\ee
where all the quantities appearing on the right hand side, 
except $\tilde{M}_\text{MSH}$, are Jordan frame quantities. 
In view 
of the consistency check presented below, however, it is 
convenient to re-express this formula in terms of  
Einstein frame quantities. Using the relations $\tilde{R} 
=  \sqrt{\phi}\, 
R $,  $ g^{ab}=\phi \, \tilde{g}^{ab} $, and $ \nabla_a \phi = 
\sqrt{ \frac{16\pi}{2\omega+3} }\, \phi \, \nabla_a 
\tilde{\phi} $, one obtains 
\begin{eqnarray}
M_\text{JF} &=& \frac{\tilde{M}_\text{MSH} }{\sqrt{\phi} }
+ \frac{2\pi \phi R^3 }{2\omega+3}  \, 
\tilde{g}^{ab}\nabla_a \tilde{\phi} \nabla_b \tilde{\phi} 
\nonumber\\
&&\nonumber\\
&\, & + \sqrt{ \frac{16\pi}{2\omega+3}}    \,
\frac{R^2}{2} \, g^{ab}\nabla_a \tilde{\phi} \nabla_b R \,.
\end{eqnarray}

Further, one notes that in spherical symmetry we have 
$\tilde{\phi}=\tilde{\phi}( 
\tilde{R})$, so $\nabla_a \tilde{\phi} = \frac{d\tilde{\phi}}{d\tilde{R}} 
\nabla_a \tilde{R}$. Then, one can use the chain rule and the definition 
$\tilde{R} = \sqrt{\phi} \, R$ to find
\be
\nabla_a R = \frac{\nabla_a  \tilde{R}}{\sqrt{\phi}} \left( 1 - 
\frac{\tilde{R}}{2} \sqrt{\frac{16\pi}{2\omega + 3}}  
\frac{d\tilde{\phi}}{d \tilde{R}} \, \right) \,.
\ee
Therefore, the last term in the definition of the mass transforms as
\be
\tilde{g}^{ab} \nabla_a \tilde{R} \, \nabla_b R = \frac{\nabla_a \tilde{R} 
\nabla_b \tilde{R}}{\sqrt{\phi}} \left( 1 - \frac{\tilde{R}}{2} 
\sqrt{\frac{16\pi}{2\omega + 3}} \frac{d\tilde{\phi}}{d \tilde{R}} \right) 
\,.
\ee
Putting these pieces together finally gives 
\begin{eqnarray}
M_\text{JF} &=& \frac{\tilde{M}_\text{MSH} }{\sqrt{\phi}}
+ \frac{2\tilde{R}^2}{\sqrt{\phi}} \sqrt{\frac{\pi}{2\omega+3}} \left( 
\frac{d\tilde{\phi}}{d\tilde{R}} \right) \tilde{g}^{ab} \nabla_a \tilde{R} 
\, \nabla_b \tilde{R} \nonumber\\
&&\nonumber\\
&\, &\cdot  \left[ 1 - \sqrt{\frac{\pi}{2\omega+3}} 
\left( \frac{d\tilde{\phi}}{d\tilde{R}} \right) \tilde{R} \right] \,. 
\label{Mspherical}
\end{eqnarray}

\section{Quasilocal mass and multipole expansion}
\setcounter{equation}{0}

A minimal requirement is that, for cylindrically symmetric and stationary  
geometries, when the mass $M_\text{JF}$ is evaluated at spatial infinity 
$ R \rightarrow + \infty$, it reduces to the monopole coefficient in a 
multipole expansion of the gravitational field. A multipole expansion 
for axisymmetric and asymptotically flat 
spacetimes in scalar-tensor gravity has been given recently by 
Pappas and Sotiriou \cite{PappasSotiriou}. Their goal was  
to express observable quantities (such as the 
frequency of quasi-periodic 
oscillations and the frequency emitted near the 
inner boundary of accretion discs orbiting black 
holes in low-mass X-ray binaries) in terms of the multipole 
moments of spacetime, which is useful in tests of gravity. 
The general multipole formalism for scalar-tensor gravity  
is rather involved, but we can perform a simple check of 
Eq.~(\ref{Mspherical}) on a well-known and very general class of 
spacetimes  (used 
also in \cite{PappasSotiriou} as a consistency check). This is the 
Just static spherically symmetric, asymptotically flat 
solution of scalar-tensor gravity \cite{Just, DamourFarese}. This 
2-parameter class of solutions is the most general static, spherically 
symmetric, asymptotically flat solution of the vacuum Brans-Dicke field 
equations with vanishing mass or potential ($V(\phi)=0$) 
which is not a black hole\footnote{The static spherical asymptotically flat 
black hole can only be the Schwarzschild black hole \cite{Hawking}, 
for which the Misner-Sharp-Hernandez 
mass trivially coincides with the Schwarzschild mass everywhere outside 
the horizon.} \cite{Bronnikov, 
FaraoniHammadCardiniGobeil}. When written in its Einstein frame representation, 
this solution is nothing but the Fisher solution of GR with 
a minimally coupled, free scalar field\footnote{This 
geometry has been rediscovered many times and is known by 
various names (Fisher-Janis-Newman-Winicour-Buchdahl-Wyman  
solution).} \cite{Fisher, Wyman}. It is the most general static, 
spherical, and asymptotically flat solution of 
the Einstein equations with a free scalar field \cite{Wyman}.
 The Einstein frame line element and scalar 
field are 
\begin{eqnarray}
&&d\tilde{s}^2= -V^{\nu}(r)dt^2 +V^{-\nu}(r)dr^2 
+V^{1-\nu}(r)r^2 d\Omega_{(2)}^2 \,,\nonumber\\
&&\\
&& \tilde{\phi}(r)=\phi_0 \ln V(r)+\phi_1 \,,\\
&&\nonumber\\
&& V(r) = 1-\frac{2\mu}{r} \,,
\end{eqnarray}
where $d\Omega_{(2)}^2=d\theta^2 +\sin^2\theta \, d\varphi^2$ is 
the line element on the unit 2-sphere, $\phi_0$, $\phi_1$, $\mu $, and $\nu$ are 
constants and, 
because of asymptotic flatness, we set $\phi_1=1$. 
In the notation 
of Ref.~\cite{PappasSotiriou} the  
exponent satisfies $\nu=m/l$, where $m$ and $l$ are length scales, 
and $\mu=l$. The areal radius of the Just geometry is clearly 
\be
\tilde{R}=r\left( 1-\frac{2\mu}{r} \right)^{\frac{1-\nu}{2} } \,.
\ee 
By applying the 
GR definition~(\ref{MSHdefinition}) in the 
Einstein frame, one obtains the Misner-Sharp-Hernandez mass of the Fisher 
spacetime 
\be
\tilde{M}_\text{MSH}=\frac{ \tilde{R} }{2} \left\{ 
1-\frac{1}{V(r)} 
\left[ 1-\frac{(\nu+1) \mu}{r} \right]^2 \right\} \,.
\ee
Since the multipoles have to be evaluated at spatial 
infinity, which corresponds  to the limits $r, R, \tilde{R} \rightarrow 
+\infty$, one finds  
\be
 \tilde{M}_\text{MSH}  \simeq  \nu\mu \,.
\ee 
Then, using the relations \cite{PappasSotiriou} $\nu=m/l$ and $\mu=l$, one 
obtains $ \tilde{M}_\text{MSH}=m$.

We can now compute the scalar-tensor mass in the Jordan 
frame for this very general geometry using 
Eq.~(\ref{Mspherical}), and then take the limit to spatial  
infinity. First, we compute the following relation

\begin{eqnarray}
\frac{d\tilde{\phi}}{d\tilde{R}}&=&\frac{d\tilde{\phi}}{dr} \, 
\left(\frac{d\tilde{R}}{dr}\right)^{-1} \nonumber\\
&&\nonumber\\
&=&  \frac{d}{dr}\left[ \phi_0 \ln V+\phi_1\right] \left\{ V^{-\, 
\frac{(\nu+1)}{2} } \left[ 1-\frac{(\nu+1)\mu}{r} \right] \right\}^{-1} 
\nonumber\\
&&\nonumber\\
&=& \frac{2\phi_0 \mu V^{\frac{\nu-1}{2}} }{r^2 \left[ 
1-\frac{(\nu+1)\mu}{r} \right] }\,.
\end{eqnarray}

Then, one obtains the following for the quasilocal mass

\begin{eqnarray}
M_\text{JF} &=&  \frac{m}{\sqrt{\phi}}
+ \frac{4\tilde{R}^2}{\sqrt{\phi}} \sqrt{\frac{\pi}{2\omega+3}}  
\frac{\phi_0 \mu V^{\frac{\nu-3}{2}}}{r^2} \nonumber\\
&&\nonumber\\ 
&\, & \cdot \left[ 1 
- \frac{\mu (1+\nu)}{r} - \sqrt{\frac{\pi}{2\omega+3}} \frac{2 \phi_0 
\mu V^{\frac{\nu-1}{2}} \tilde{R}}{r^2} \right] \,.\nonumber\\
&&
\end{eqnarray}

In the limit $r\rightarrow +\infty$, we have $\phi\rightarrow 
1$, $V(r) \rightarrow 1$, and  the quasilocal mass seen from spatial 
infinity is
\be
M_\text{JF}^\text{(Just)} \simeq m+\sqrt{ 
\frac{16\pi}{2\omega+3}} \, \phi_0 l \,. \label{eq:acci}
\ee

Let us now compare our prescription~(\ref{eq:acci}) with the monopole 
coefficient. The scalar-tensor multipole expansion 
of Ref.~\cite{PappasSotiriou} gives the Jordan frame monopole coefficient
\be
M_\text{JF}=m - w_A \alpha_0 \,,
\ee
and the Einstein frame scalar field is written as 
\be
\tilde{\phi}=\frac{w_A}{2l} \, \ln \left( 1- \frac{2l}{r} \right) \,,
\ee 
so $w_A=2l\phi_0$. Ref.~\cite{PappasSotiriou} uses the widespread 
notation \cite{DamourFarese} $A(\phi)=\Omega^{-1}=\phi^{-1/2}$, and $ \alpha_0 \equiv \left. d\left( \ln A 
\right) / d\tilde{\phi} \right|_{\infty}$. It is then  
straightforward to derive
\be
\alpha_0 = \left. -\frac{1}{2\phi} \, \frac{d\phi}{d\tilde{\phi}} 
\right|_{\infty}  =  
- \sqrt{ \frac{4\pi}{2\omega+3}} 
\ee
and 
\be
M_\text{JF} = m +4 l\phi_0 \sqrt{\frac{\pi}{2\omega+3}}  
\,,
\ee
which coincides with the expression~(\ref{eq:acci}) of 
$M_\text{JF}^\text{(Just)} $. The quasilocal mass introduced 
in Ref.~\cite{mySTmass}  reproduces (in the limit to spatial infinity) 
the  monopole coefficient obtained in a multipole expansion 
which was performed entirely in the Jordan frame and is 
completely independent of our considerations. 


\section{Conclusions}
\setcounter{equation}{0}

As a partial check of the reliability of the quasilocal mass prescription 
of~\cite{mySTmass} in scalar-tensor gravity, we have shown that, in a very 
general  vacuum situation, this construct reproduces (in 
the limit to spatial infinity) the  monopole coefficient provided by the  
recent multipole expansion of Ref.~\cite{PappasSotiriou}. This multipole 
formalism was derived entirely in the Jordan frame, in a way  
completely independent of our considerations. This fact inspires  
confidence in the recipe of Ref.~\cite{mySTmass} for extending the 
quasilocal  mass to scalar-tensor gravity. 

On the one hand, the extent of our check is limited because, in order to 
have a multipole expansion, one needs to restrict to stationary and 
asymptotically flat 
geometries and it is necessary to look at the physical system from spatial 
infinity. Moreover, a multipole expansion of the relativistic 
gravitational field is complicated and its implementation in scalar-tensor 
gravity is quite recent \cite{PappasSotiriou}. On the other hand, we have 
applied the quasilocal prescription to the most general static, 
spherically symmetric, and asymptotically flat solution of Brans-Dicke 
theory with  free scalar field that is not a black hole \cite{Bronnikov, 
FaraoniHammadCardiniGobeil}. All stable static, spherical, and 
asymptotically flat black hole solutions of (even more general) 
scalar-tensor theories 
reduce to the Schwarzschild black hole \cite{Hawking}, and the quasilocal 
mass is then trivial. 

Further tests of the quasilocal mass prescription of Ref.~\cite{mySTmass}  
and its application to horizon thermodynamics will be given in future 
work.

\begin{acknowledgments}

This work is supported by the Natural Sciences and Engineering Research 
Council of Canada (Grant No.~2016-03803 to V.F.) and by Bishop's 
University.

\end{acknowledgments}

\end{document}